# Tunable energy transfer between dipolar-coupled magnetic disks by stimulated vortex gyration

Hyunsung Jung,[1] Ki-Suk Lee,[1] Dae-Eun Jeong,[1] Youn-Seok Choi,[1] Young-Sang Yu,[1] Dong-Soo Han,[1] Andreas Vogel,[2] Lars Bocklage,[2] Guido Meier,[2] Mi-Young Im,[3] Peter Fischer,[3] & Sang-Koog Kim [1,3]*

[1]National Creative Research Center for Spin Dynamics & Spin-Wave Devices, and Nanospinics Laboratory, Department of Materials Science and Engineering, Seoul National University, Seoul 151-744, Republic of Korea

[2]Institut für Angewandte Physik und Zentrum für Mikrostrukturforschung, Universität Hamburg, 20355 Hamburg, Germany

[3]Center for X-ray Optics, Lawrence Berkeley National Laboratory, Berkeley CA 94720, USA

*e-mail: sangkoog@snu.ac.kr, The author recently was on sabbatical research leave at the Center for X-Ray Optics, Materials Science Division, Lawrence Berkeley National Laboratory.

A wide variety of coupled harmonic oscillators exist in nature[1]. Coupling between different oscillators allows for the possibility of mutual energy transfer between them[2-4] and the information-signal propagation[5,6]. Low-energy input signals and their transport with low-energy dissipation are the key technical factors in the design of information processing devices[7]. Here, utilizing the concept of coupled oscillators, we experimentally demonstrated a robust new mechanism for energy transfer between spatially separated dipolar-coupled magnetic disks - stimulated vortex gyration. Direct experimental evidence was obtained by time-resolved soft X-ray microscopy. The rate of energy transfer from one disk to the other was deduced from the two normal modes' frequency splitting caused by dipolar interaction. This mechanism provides the advantages of tunable energy transfer rate, low-power input signal, and low-energy dissipation for magnetic elements with negligible damping. Coupled vortex-state disks are promising candidates for information-signal processing devices that operate above room temperature.

The magnetic vortex structure is strongly stable as the ground state in nanoscale magnetic elements[8,9]. It is characterized by an in-plane curling magnetization (chirality) around and an out-of-plane magnetization (polarization) in the central (core) region. In isolated disks, applied magnetic fields or spin currents induce vortex excitations, among which a translational mode [10-14] exists in which the vortex core rotates around its equilibrium position at a characteristic eigenfrequency ($\nu_0 = \omega_0/2\pi$) [15] typically ranging from several hundred MHz to ~1 GHz. This so-called gyrotropic motion occurs because the gyroforce of the vortex core is in balance with the restoring force acting on it, both of which forces are due to the competition between the short-range exchange and long-range dipolar interactions [15,16]. The rotation sense of the vortex gyration is determined by the polarization $p$ (Ref. 11, 15) and thus is controllable by the $p$ reversal, where $p$ = +1 (-1) corresponds to the upward (downward) orientation of the magnetizations in the core. When the angular frequency of a driving force is close to $\omega_0$, the vortex-core motion and its switching are resonantly driven with low power consumption[17-19]. Such a vortex oscillator functions analogously to a simple harmonic oscillator. Recently the vortex structure is attracting keen interest owing to its potential application to nano-oscillators that emit microwave signals[16,20].

Moreover, vortex oscillators in two physically separated magnetic disks can be coupled via magnetostatic interaction [21-25]. A displaced vortex core and its motion in one disk generate dynamically rotating stray fields, consequently affecting the potential energy of the other disk, in which vortex gyration can be stimulated resonantly. The relative vortex-core displacements in both disks as well as the disk-to-disk interdistance modify the interaction strength. This behaviour is analogous to coupled harmonic oscillators such as coupled pendulums or capacitively-coupled inductor-capacitor resonators[1]. Under the free-relaxation condition of such coupled oscillators, the kinetic and potential energy of one oscillator can be transferred to the other [1].

In the present study, we identified reliably controllable dipolar-coupled vortex-core nano-oscillators by a direct experimental observation of energy transfer and the collective normal modes. This mechanism is a robust means of tunable energy transfer and information-signal transport between physically separated magnetic disks, and provides for the advantages of low-energy-dissipation signal transmission and low-power input signals.

Figure 1 shows the sample, which contains several two-disk pairs, each of which consists of two Permalloy (Py: $Ni_{80}Fe_{20}$) disks of the same dimensions. Note that both disks are positioned on the *x* axis, which is referred to as the bonding axis. The dipolar interaction between the two disks along this axis breaks the radial symmetry of the potential well of both isolated disks. Applying a pulse current of 90 ns width and 2.5 ns rise and fall time along a single-strip Cu electrode (along the *y* axis) placed on only one disk (here denoted as disk 1), we shifted the core position in disk 1 to ~165 nm in the +*y* direction from the center position $(x_1, y_1) = (0,0)$, and then allowed it to relax after turning off the field generated by the current pulse. During this free relaxation, the dynamics of the vortex gyrations in both disks of each pair were observed simultaneously, using spatiotemporal-resolved full-field magnetic transmission soft X-ray microscopy (MTXM). The magnetic contrast was provided via X-ray magnetic circular dichroism (XMCD) at the Fe $L_3$ edge (~707 eV). Measurements were made on the basis of a stroboscopic pump-and-probe technique [25-27] (see Supplementary A for details). Figure 1b shows the two pairs studied, that is, $d_{int}/(2R) = 1.05$ and 1.10, and includes simulation-perspective images of the initial vortex ground states in both disks. For the other pairs having larger $d_{int}/(2R)$ values, the gyrations in disk 2 were difficult to observe, due to the small deviation of the vortex core from the center position, caused by the rapidly reduced interaction strength.

Figure 2 shows the experimental data on the vortex gyrations in the two disks in both pairs: $d_{int}/(2R) = 1.05$ and 1.10 in (a) and (b), respectively (see also Supplementary Movies. 1 and 2). Representative serial snapshot images of the XMCD contrasts are shown in each top panel. The trajectory curves of the motions of both vortex-core position vectors, $\mathbf{X}_1 = (x_1, y_1)$ and $\mathbf{X}_2 = (x_2, y_2)$, are plotted in the lower-right panels together with, in the lower-left panels, their oscillatory $x$ and $y$ components. Since a quasi-static local magnetic field (a pulse of sufficient length: here, 90 ns) was applied only to disk 1, the core position in that disk was shifted to $(x_1, y_1) = (0, +165$ nm) for $[p_1, C_1] = [+1, +1]$ and to $(x_1, y_1) = (0, -165$ nm) for $[p_1, C_1] = [+1, -1]$. The opposite sign of the $y$ displacement in disk 1 between $d_{int}/(2R) = 1.05$ and 1.10 is attributable to the opposite chirality of disk 1 in the two pairs. Once the pulse field was turned off (at $t = 0$, the reference), the vortex core in disk 1 began to gyrate with decreasing oscillation amplitude starting from $(x_1, y_1) = (0, +165$ nm$)$, whereas the vortex core in disk 2 began to gyrate with increasing oscillation amplitude starting from $(x_2, y_2) = (0, 0)$. The vortex core in disk 1 gyrated counter-clockwise, indicating the upward core ($p_1 = +1$) orientation, whereas the vortex core in disk 2 gyrated clockwise, corresponding to the downward core ($p_1 = -1$) orientation (see Fig. 2a). Figure 2b shows similar characteristic oscillations of the given vortex state configuration. Micromagnetic simulation results (see Methods) are in excellent agreements with the experimental one, except for small discrepancies of their frequencies and absolute amplitudes. Experimental results show smaller oscillation amplitudes and frequencies, which might be attributed to the imperfection of the sample and weaker interaction between both disks due to a possible reduction in the saturation magnetization of the sample.

The most important finding here is that the vortex-core gyration in disk 2 is stimulated by the vortex-core gyration in disk 1, not by any external driving force or pulse field, but rather through the magnetostatic interaction between the two disks. For the case of $d_{int}/(2R) = 1.05$, the decreasing amplitude in disk 1 started to increase again at the time $t = 26$ ns, whereas the increasing amplitude in disk 2 began to decrease (see Fig. 2a). For $d_{int}/(2R) = 1.10$, the amplitude decay in disk1 occurs at a longer time, $t = \sim 45$ ns (see Fig. 2b). Since the magnetic potential energy of a vortex depends on the displacement of the vortex core from the disk center[15], energy transfer between dipolar-coupled vortex-state disks can be observed directly, via the variations of the displacements of both vortex cores, that is, $|\mathbf{X}_1|$ and $|\mathbf{X}_2|$, as shown in Fig. 3. Such a crossover in the vortex-core gyration modulation envelope between disks 1 and 2 confirms that the energy stored via the displaced core by an input-signal pulse field is transferred from disk 1 to disk 2 through the robust mechanism of stimulated vortex gyration in dipolar-coupled vortex oscillators. The high efficiency of this energy transfer mechanism is evidenced by the maximal oscillating amplitudes of the exciting and the excited vortex oscillators.

The total energy of undamped coupled oscillators, which is the sum of the first and second oscillator energies and the interaction energy between the oscillators, is constant. Magnetic vortices typically reside in potentials, which give rise to corresponding eigenfrequencies up to ~ 1 GHz. However, in principle these frequencies can be modified in virtually any desired frequency range, which is an important and advantageous aspect of the application of coupled vortex oscillators. As in every coupled harmonic oscillator system, the original resonance frequency of the non-interacting oscillator is either enhanced or reduced by the coupling[1]. Generally a symmetric and an antisymmetric mode having a lower and a higher frequency compared with the original eigenfrequency in uncoupled disks, respectively, appear. The frequency splitting between the symmetric and antisymmetric modes have a direct correlation with the interaction energy, and can be deduced from the beating pattern of coupled oscillations (see Supplementary B). It is clear that such a modulation envelope crossover between disks 1 and 2, as demonstrated in Figs. 2 and 3, is the result of the superposition of two equally weighted normal modes of different frequencies excited in a given dipolar-coupled oscillator system. The beating frequency $\Delta\omega$ of the modulation envelope functions in disk 1 and disk 2 can be obtained by fitting the data of $|\mathbf{X}_1|$ and $|\mathbf{X}_2|$ to two orthogonal sinusoidal functions with a damping term, $|\mathbf{X}_1| = |A\cos(\Delta\omega t/2)|\exp(-\beta t)$ and $|\mathbf{X}_2| = |A\sin(\Delta\omega t/2)|\exp(-\beta t)$: $\Delta\omega/2\pi$ = 20 MHz and 12 MHz for $d_{\text{int}}/(2R)$ = 1.05 and 1.10, respectively. These results agree well on the

fact that this angular frequency splitting $\Delta\omega$ takes place according to interaction between oscillators in a coupled system (see Supplementary C) and the magnitude of $\Delta\omega$ varies with the interdistance between neighbouring oscillators.

To directly obtain the value of $\Delta\omega$ from the experimentally observed vortex gyrations in the coupled system, we transformed the *x* and *y* component oscillations in each disk into two normal-mode oscillations on the basis of normal-mode coordinates $(N1_X, N1_Y) = (x_1+x_2, y_1-y_2)$ and $(N2_X, N2_Y) = (x_1-x_2, y_1+y_2)$ for the case of $p_1p_2 = -1$. In the normal-mode representation, the experimentally observed vortex gyrations in real coordinates (*x*, *y*) were decoupled into two normal modes having corresponding single dominant frequencies, as shown in Fig. 4 (see also the two normal modes for a different value, $d_{int}/(2R) = 1.10$, in Supplementary D). Figures 4a reveals that the two normal modes were excited almost equally and that their amplitudes were damped monotonically, as in free-relaxation vortex-core gyrations in uncoupled disks. According to their FFT spectra in Fig. 4b, the lower and higher eigenfrequencies for the two normal modes were 145 ± 10 and 165 ± 10 MHz, and thus, their frequency splitting was $\Delta\omega/2\pi$ = 20 MHz for $d_{int}/2R$ = 1.05, which is in excellent agreement with the values obtained from the fit to the modulation envelope functions of $|\mathbf{X}_1|$ and $|\mathbf{X}_2|$ with the corresponding damping (Fig. 3).

The energy exchange rate $\tau_{ex}$ between oscillators is technologically important from the information-signal transport point of view. It is determined by the coupling strength. Let $\tau_{ex}$ be the time period required for transferring the potential energy stored in disk 1 (here, due to the initial vortex-core shift) completely to disk 2. $\tau_{ex}$ can be estimated from the frequency splitting determined from the modulation envelopes of $|\mathbf{X}_1|$ and $|\mathbf{X}_2|$, (see Supplementary E) and consequently, is given as half of the modulation envelope period, $\frac{1}{2}(2\pi/\Delta\omega)$, so that $\tau_{ex}$ = 26 ± 3 ns and 45 ± 6 ns for $d_{int}/(2R)$ = 1.05 and 1.10, respectively. Since the strength of the interaction between two vortices determines the frequency splitting of dipolar-coupled vortices, a smaller value of $\tau_{ex}$ can be obtained by increasing the strength of the dipolar interaction. For any given material and vortex state, it is known that the dipolar interaction between two vortices depends very strongly on the interdistance between them[21,22]. This was confirmed by the present experiments. Also, the micromagnetic simulation data (see Supplementary F) clearly show that the value of $\tau_{ex}$ varies with the interdistance as $\tau_{ex} \sim \tau_{ex} \sim (d_{int}/2R)^{-3.91\pm0.07}$.

Accordingly, the energy transfer rate between vortex-state disks is tunable varying their interdistance.

It is worth noting that the relative motion of the vortex-core gyration of two vortices also affects the strength of their interactions: the dipolar interaction of vortex-core gyrations having opposite rotation senses due to opposite polarities is much stronger than that of gyrations having the same rotation sense due to the same polarity. From the simulation (see Supplementary G), $\tau_{ex}$ for the opposite polarity is ~ 2 times faster than that for the same polarity. Since the magnetostatic interaction strength depends directly on the stray field of each disk, thus, the larger the saturation magnetization is, the stronger will be the interaction strength, yielding faster $\tau_{ex}$ and hence signal transport.

To conclude, we experimentally observed, by time-resolved MTXM, energy and information-signal transfer via stimulated vortex gyration through dipolar interaction between separate magnetic disks. This robust new mechanism for energy transfer provides the advantages of a fast and tunable energy transfer rate that is a function of disk interdistance and interaction strength. Even lower energy dissipation is possible for magnetic elements with negligible damping. Vortex gyration also can be achieved with low-power consumption through the resonant vortex excitation. This finding opens a new avenue to the development of novel fast, small, versatile and energy efficient information processing devices.

## Methods

**Sample preparation**. The sample shown in Fig. 1 was prepared on a 100 nm-thick silicon nitride membrane by electron-beam lithography, thermal evaporation, and lift-off processing. The two disks in each pair have a different center-to-center interdistance, such that $d_{int}/2R$ = 1.05, 1.1, 1.15, and 1.2, and each pair is separated at a sufficiently large distance, 5.6 $\mu$m, from neighbouring pairs. A single-strip Cu electrode leads to strong local fields, as shown in Supplementary G, so that local excitations of the vortex, in each pair, are possible only in disk 1. Note that we confirmed that such local fields do not allow for vortex excitations in disk 2 positioned sufficiently far from the electrode stripline, in any of the $d_{int}/2R$ cases studied here [28]. The eigenfrequency of the vortex gyrations in isolated Py disks, by ferromagnetic resonance measurements performed on an array of Py disk pairs of the same dimensions and $d_{int}/2R$ = 1.56, was determined to be around 157 $\pm$ 3 MHz (Ref. 29).

**Micromagnetic simulation.** We performed micromagnetic simulations of coupled vortex gyrations under free relaxation for two Py disks of the same dimensions as those of the sample for $d_\text{int}/(2R)$ = 1.05 and 1.10. To mimic the experimental conditions, we used the same initial vortex-state configurations as those of the real sample. The material parameters used in this simulation are as follows: the exchange stiffness $A_\text{ex}$ =13 pJ/m, the saturation magnetization $M_s$=7.2 × 10$^5$ A/m, and a zero magnetic anisotropy constant. The cell size was 4×4×50 nm$^3$ with the damping constant α= 0.01. We used the OOMMF code that utilizes the Landau-Lifshitz-Gilbert equation of motion[30].

# References

1. Thornton, S. T. & Marion, J. B. *Classical dynamics of particles and systems, Fifth Ed.* (Thomson, 2004).

2. Serway, R. A. *Principles of Physics, Second Ed.* (Saunders college publishing, 1998).

3. Woutersen, S. & Bakker, H. J. Resonant intermolecular transfer of vibrational energy in liquid water. *Nature* **402**, 507-509 (1999).

4. Sisourat, N. *et al.* Ultralong-range energy transfer by interatomic Coulombic decay in an extreme quantum system. *Nature Phys.* **6**, 508-511 (2010).

5. Cowburn, R. P. & Welland, M. E. Room temperature magnetic quantum cellular automata. *Science* **287**, 1466-1468 (2000).

6. Carlton, D. B., Emley, N. C., Tuchfeld, E. & Bokor, J. Simulation studies of nanomagnet-based logic architecture. *Nano Lett.* **8**, 4173-4178 (2008).

7. Bandyopadhyay, S. & Cahay, M. Electron spin for classical information processing: a brief survey of spin-based logic devices, gates and circuits. *Nanotechnology* **20**, 412001 (2009).

8. Shinjo, T., Okuno, T., Hassdorf, R., Shigeto, K. & Ono, T. Magnetic vortex core observation in circular dots of Permalloy. *Science* **289**, 930-932 (2000).

9. Wachowiak, A. *et al.* Direct observation of internal spin structure of magnetic vortex cores. *Science* **298**, 577-580 (2002).

10. Park, J. *et al.* Imaging of spin dynamics in closure domain and vortex structures. *Phys. Rev. B* **67**, 020403 (2003).

11. Choe, S.B. *et al.* Vortex core-driven magnetization dynamics. *Science* **304**, 420-422 (2004).

12. Kasai, S. *et al.* Current-driven resonant excitation of magnetic vortices. *Phys. Rev. Lett.* **97**, 107204 (2006).

13. Bolte, M. *et al.* Time-resolved x-ray microscopy of spin-torque-induced magnetic vortex gyration. *Phys. Rev. Lett.* **100**, 176601 (2008).

14. Lee, K.-S. & Kim, S.-K. Two circular-rotational eigenmodes and their giant resonance asymmetry in vortex gyrotropic motions in soft magnetic nanodots. *Phys. Rev. B* **78**, 014405 (2008).

15. Guslienko, K. Y. *et al.* Eigenfrequencies of vortex state excitations in magnetic submicron-size disks. *J. Appl. Phys.* **91**, 8037-8039 (2002).

16. Dussaux, A. *et al.* Large microwave generation from current-driven magnetic vortex oscillators in magnetic tunnel junctions. *Nat. Commun.* 1**:**8*,* doi: 10.1038/ncomms1006 (2010).

17. Waeyenberge, B. *et al.* Magnetic vortex core reversal by excitation with short bursts of an alternating field. *Nature* **444**, 461-464 (2006).

18. Yamada, K. *et al.* Electrical switching of the vortex core in a magnetic disk. *Nature Mater.* **6**, 269-273 (2007).

19. Kim, S.-K., Lee, K.-S., Yu, Y.-S. & Choi, Y.-S. Reliable low-power control of ultrafast vortex-core switching with the selectivity in an array of vortex states by in-plane circular-rotational magnetic fields and spin-polarized currents. *Appl. Phys. Lett*. **92**, 022509 (2008).

20. Pribiag, V. S. *et al.* Magnetic vortex oscillator driven by d.c. spin-polarized current, *Nature Phys.* **3**, 498-503 (2007).

21. Shibata, J., Shigeto, K. & Otani, Y. Dynamics of magnetostatically coupled vortices in magnetic nanodisks. *Phys. Rev. B* **67**, 224404 (2003).

22. Vogel, A., Drews, A., Kamionka, T., Bolte, M. & Meier, G. Influence of dipolar Interaction on vortex dynamics in arrays of ferromagnetic disks. *Phys. Rev. Lett.* **105**, 037201 (2010).

23. Barman, S., Barman, A. & Otani, Y. Dynamics of 1-D Chains of Magnetic Vortices in Response to Local and Global Excitations. *IEEE Trans. Magn.* **79**, 1342-1345 (2010).

24. Barman, A., Barman, S., Kimura, T., Fukuma, Y. & Otani, Y. Gyration mode splitting in magnetostatically coupled magnetic vortices in an array. *J. Phys. D: Appl. Phys.* **43**, 422001 (2010).

25. Jung, H. *et al.* Observation of coupled vortex gyrations by 70-ps-time- and 20-nm-space-resolved full-field magnetic transmission soft x-ray microscopy. *Appl. Phys. Lett.* (in press).

26. Bocklage, L. *et al.* Time-resolved imaging of current-induced domain-wall oscillations, *Phys. Rev. B* **78**, 180405(R) (2008).

27. Fischer, P. Soft X-ray microscopy –a powerful analytical tool to image magnetism down to fundamental length and time scales. *AAPPS bulletin* **18**(6), 12-17 (2008).

28. Referencing a different sample that had no Py disk beneath the long Cu electrode, we were able to confirm that the local field did not excite either disk far from the stripline electrode.

29. For the ferromagnetic resonance measurement, an array of Py disk pairs was positioned under a 7 μm-wide stripline. The disks were of the same dimensions ($2R$ = 2.4 μm, $L$ = 50 nm), but a larger value of $d_{int}/(2R)$=1.56 was applied, to prevent interaction.

30. We used the OOMMF code. See http://math.nist.gov/oommf.

# Acknowledgments

This research was supported by the Basic Science Research Program through the National Research Foundation of Korea (NRF), funded by the Ministry of Education, Science, and Technology (grant No. 20100000706). The operation of the microscope was supported by the Director, Office of Science, Office of Basic Energy Sciences, Materials Sciences and Engineering Division, of the U.S. Department of Energy. Financial support of the Deutsche Forschungsgemeinschaft via the SFB 668 "Magnetismus vom Einzelatom zur Nanostruktur" and via the Graduiertenkolleg 1286 "Functional Metal-Semiconductor Hybrid Systems" is gratefully acknowledged, as is that of the City of Hamburg via Cluster of Excellence "Nano-Spintronics."

**Figure Legends**

**Figure 1| Sample geometry of pairs of two vortex-state disks and initial ground states**. **a**, Scanning electron microscopy image of sample containing several two-disk pairs. Each pair contains two Permalloy (Py: $Ni_{80}Fe_{20}$) disks of the same dimensions, as indicated in the inset. The center-to-center distance normalized by diameter $d_{int}/(2R)$ varies, such that $d_{int}/(2R)$ = 1.05, 1.1, 1.15, 1.2. Each pair is separated 5.6 $\mu$m from the neighbouring pairs. The insets show schematic illustrations of the sample with the indicated dimensions and the employed current pulse of 90 ns width, 2.5 ns fall-and-rise time, and $7.0\times10^6$ A/cm$^2$ current density. **b**, Fe $L_3$ edge XMCD images of both disks for two pairs, $d_{int}/(2R)$ = 1.05 and 1.10, where initial ground states are represented by perspective-view simulation results, as indicated.

**Figure 2| MTXM observation of vortex-core gyrations in dipolar-coupled Py disks for $d_{int}/(2R)$ = 1.05 in (a) and 1.10 in (b), compared with the corresponding simulation results.** In each of a and b, the first row, are serial

snapshot XMCD images of temporal evolution of vortex-core gyrations, starting from $t$ = 8.8 ns (①) and ending at $t$ = 19.7 ns (⑧) for a), and starting from $t$ = 12.5 ns (①) and ending at $t$ = 24.2 ns (⑧) for b). In each of a) and b), the second (disk 1) and fourth (disk 2) row are experimental results of the $x$ and $y$ components of vortex-core positions (left) from center position ($x$, $y$)=(0,0), and constructed vortex-core trajectories (right). The third and fifth row indicate the corresponding simulation results.

**Figure 3 | Comparison of vortex-core displacement variations versus time in both disks for $d_{int}/(2R)$ = 1.05 and 1.10.** The open and closed circles represent the experimental results for disks 1 and 2, respectively. The thick solid curves are the results of the fits to $|\mathbf{X}_1| = |A\cos(\Delta\omega t/2)|\exp(-\beta t)$ and $|\mathbf{X}_2| = |A\sin(\Delta\omega t/2)|\exp(-\beta t)$, where amplitude $A = 141 \pm 12$ nm, damping coefficient $\beta = 2\pi \times (4.2 \pm 0.7)$ MHz, $\Delta\omega = 2\pi \times (19.3 \pm 2)$ MHz for $d_{int}/(2R)$ = 1.05 and $A = 166 \pm 15$ nm, $\beta = 2\pi \times (3 \pm 0.7)$ MHz, and $\Delta\omega = 2\pi \times (11.2 \pm 1.5)$ MHz for $d_{int}/(2R)$ = 1.10. The vertical lines correspond to the rate of energy transfer between the two disks: $\tau_{ex} = \frac{1}{2}(2\pi/\Delta\omega)$.

**Figure 4 | Normal-mode representations of vortex-core gyrations in dipolar-coupled oscillators for $d_{int}/(2R)$ = 1.05.**

**a**, Oscillatory core motions and **b**, dominant frequency spectra. The two normal-mode coordinates are represented as ($N1_X = x_1+x_2$, $N1_Y = y_1-y_2$) and ($N2_X = x_1-x_2$, $N2_Y = y_1+y_2$), corresponding to their trajectory curves and FFT spectra.

**Figure 1.**

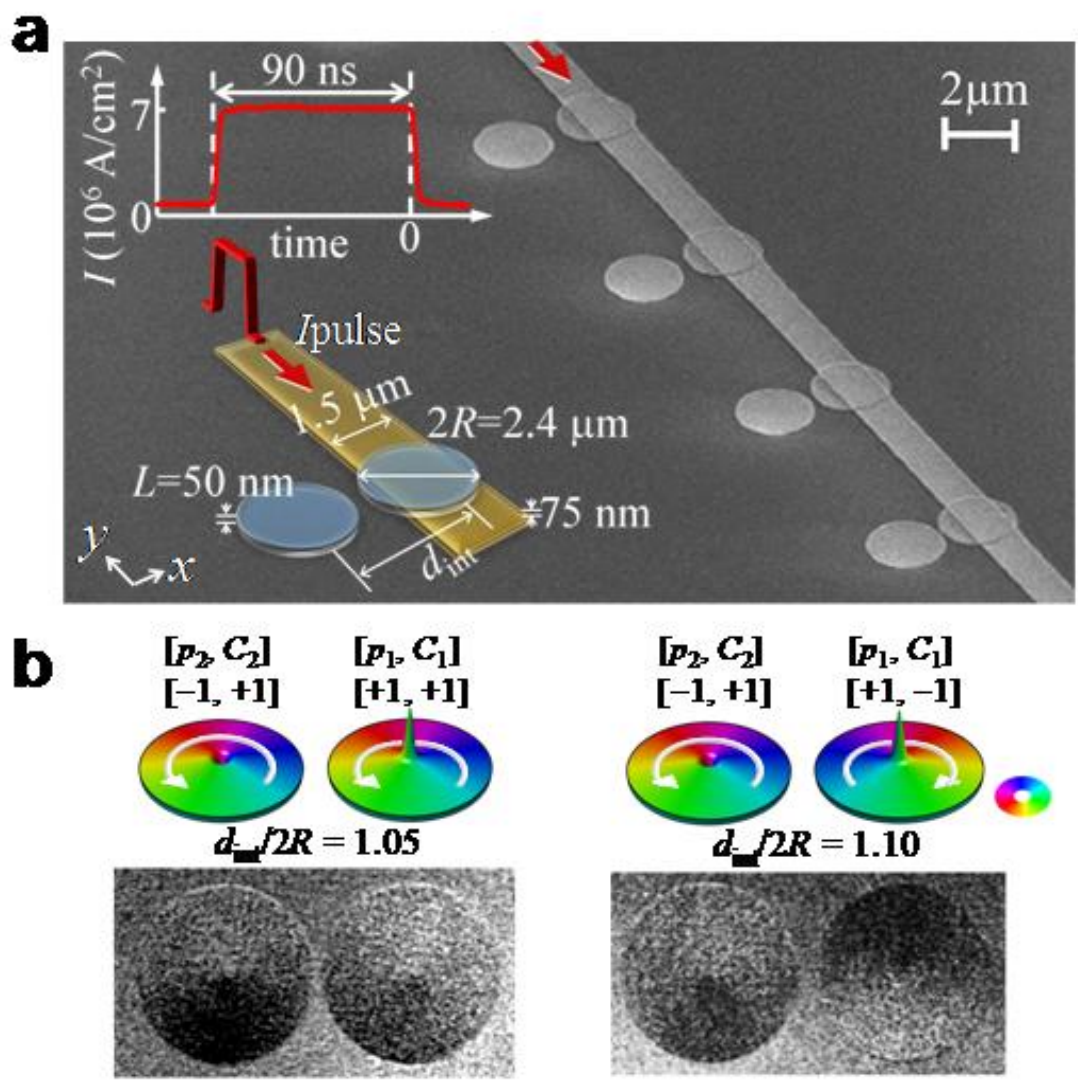

**Figure 2.**

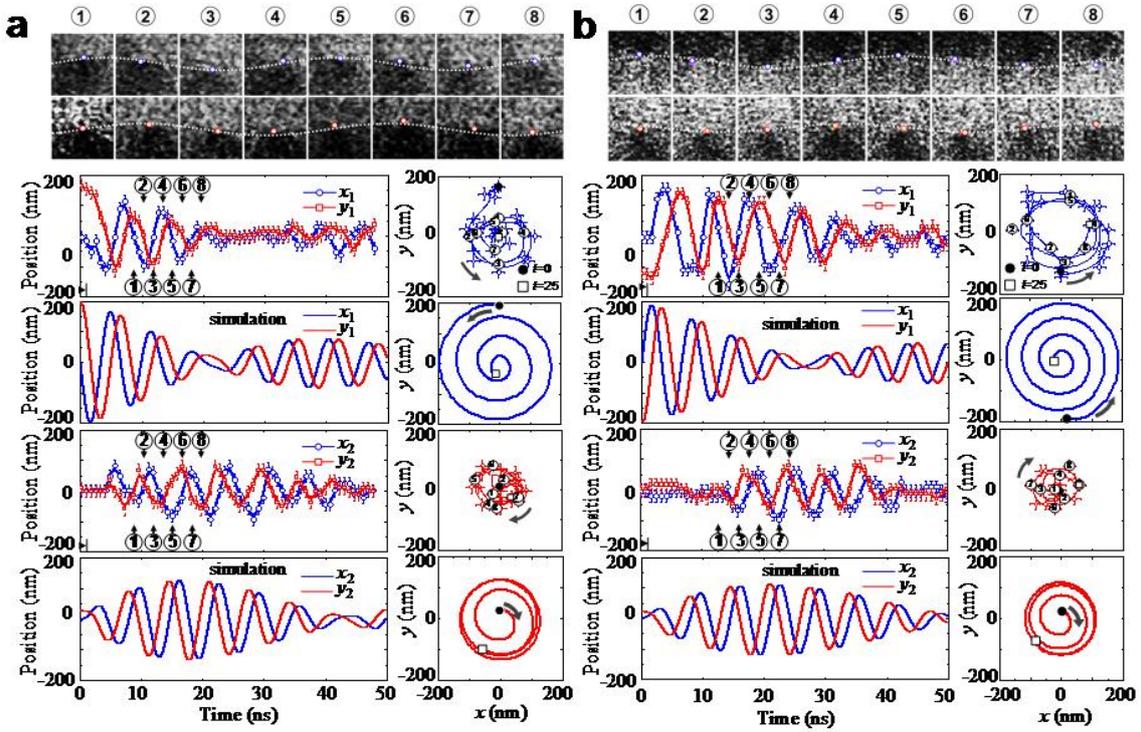

**Figure 3,**

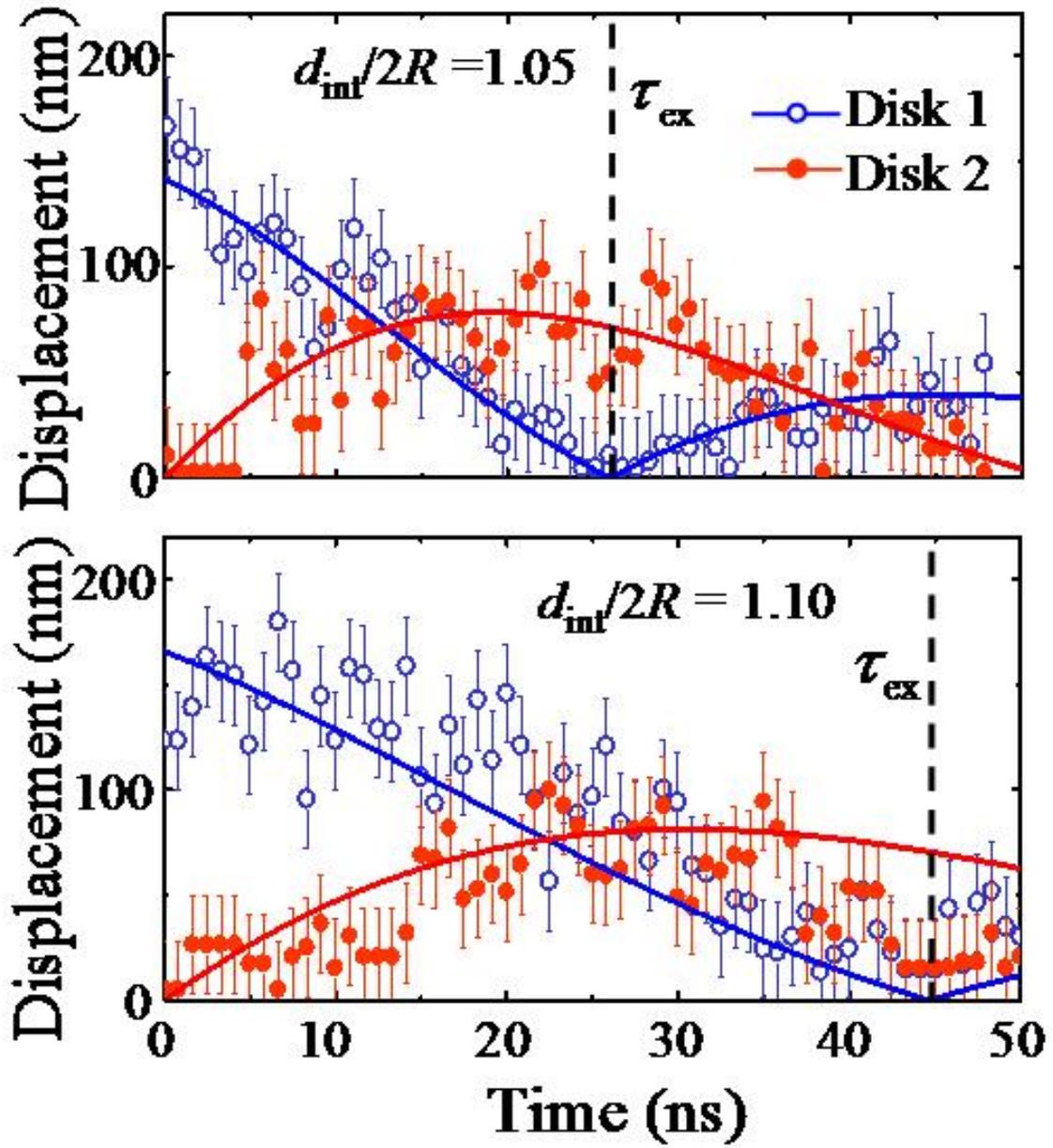

**Figure 4.**

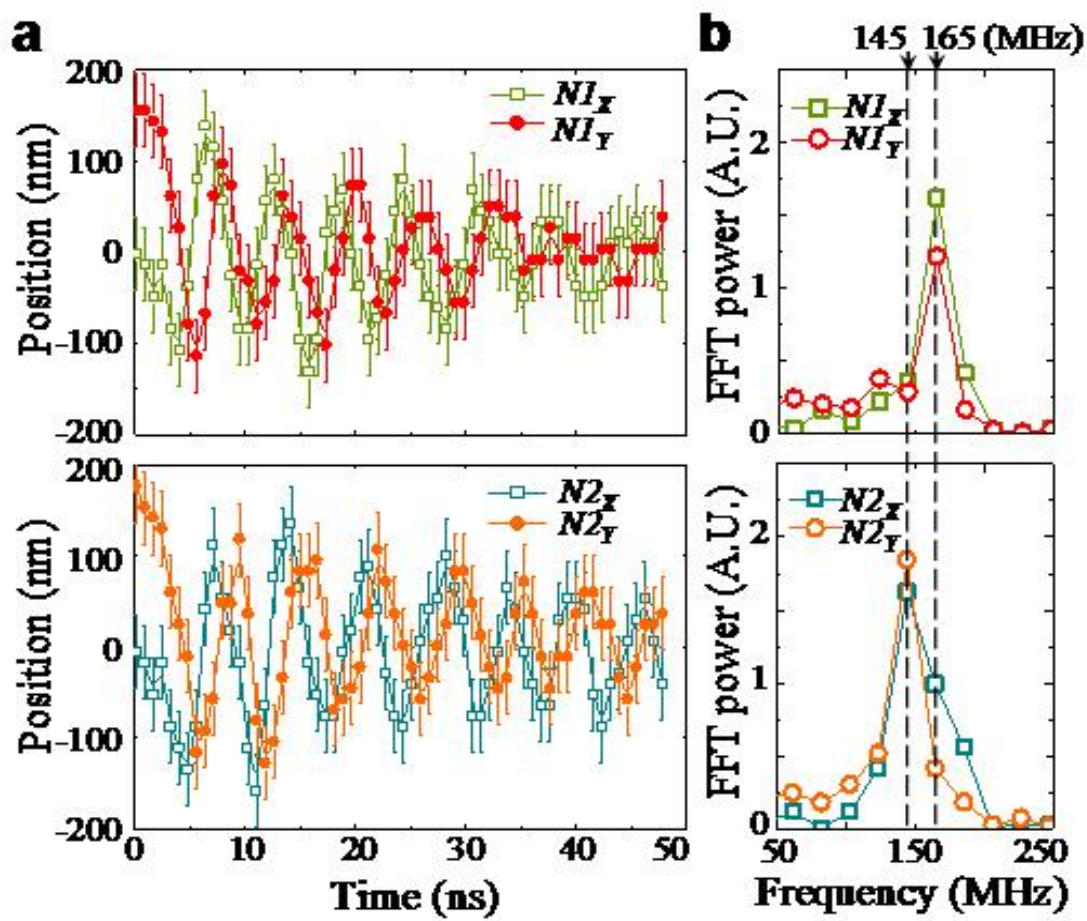